\documentstyle[epsfig]{aipproc}

\begin{document}
\title{X-ray Observations of Neutron Star Binaries: Evidence for Millisecond
Spins}

\author{Tod E. Strohmayer}
\address{Laboratory for High Energy Astrophysics \\ NASA's Goddard Space Flight
Center \\ Mail Code 662, Greenbelt, MD 20771}

\maketitle

\begin{abstract}

High amplitude X-ray brightness oscillations during thermonuclear
X-ray bursts were discovered with the {\it Rossi X-ray Timing Explorer} (RXTE)
in early 1996. Spectral and timing evidence strongly supports the 
conclusion that these oscillations are caused by rotational modulation
of the burst emission and that they reveal the spin frequency of
neutron stars in low mass X-ray binaries (LMXB), a long sought goal of X-ray
astronomy. I will briefly review the status of our knowledge of these 
oscillations. So far 10 neutron star systems have been observed to 
produce burst oscillations, interestingly, the observed frequencies
cluster in a fairly narrow range from $\sim 300 - 600$ Hz, well below the
break-up frequency for most modern neutron star equations of state (EOS). 
This has led to suggestions that their spin frequencies may be limited by 
the loss of angular momentum due to gravitational wave emission. 
Connections with gravity wave rotational instabilities will be briefly 
described.

\end{abstract}

\section*{Introduction}

X-ray binaries are potentially among the most interesting sources of 
gravitational wave emission which current and future gravity wave detectors 
will attempt to study. The high frequency gravity wave signal produced during 
binary inspiral and ring down of black hole and neutron star binaries contains 
detailed information on the properties of the compact object as well as the 
structure of spacetime in its vicinity (see Lee; Faber \& Rasio; these 
proceedings). These objects will be 
prime targets for ground based detectors such as LIGO which because of seismic
noise are only sensitive in the high frequency range above $\sim 100$ Hz 
(see Barish, these proceedings).
\par Neutron stars are compelling targets of investigation because of the 
extreme physical conditions which exist in their interiors and immediate 
environs. For example, the gravity wave signals produced by inspiral of a 
neutron star depend on the equation of state (EOS) at supranuclear density, a 
quantity which is still not precisely constrained by currently available 
astrophysics and nuclear physics data (see for example Heiselberg \& 
Hjorth-Jensen 1999). Moreover, fundamental 
properties of the star, such as its mass, can be extracted if the gravity wave 
signal can be measured. Thus gravity wave astronomy can in principle provide 
new probes of fundamental physics as well as advancing neutron star 
astrophysics.
\par Radio observations provided the first indications that some neutron stars 
are spinning with periods approaching 1.5 ms (Backer et al. 1982). These 
rapidly rotating neutron stars are observed as  either isolated or binary 
radio pulsars. Binary evolution models indicate that neutron stars accreting 
mass from a companion can be spun up, or `recycled', to millisecond periods 
(see for example Webbink, Rappaport \& Savonije 1983). 
This formation mechanism likely accounts for a substantial fraction of the 
observed population of millisecond radio pulsars, however, other formation 
scenarios have also been proposed (van den Heuvel \& Bitzaraki 1995; van 
Paradijs et al. 1997). In recent years direct evidence linking 
the formation of rapidly rotating neutron stars to accreting X-ray binaries has
been provided by data from the {\it Rossi X-ray Timing Explorer} (RXTE). The 
first evidence came from the discovery of high amplitude, nearly coherent X-ray
brightness oscillations (so called `burst oscillations') during thermonuclear 
flashes from several neutron star binaries (see Strohmayer 2000 for a recent 
review). These oscillations likely result from spin modulation of either one or
a pair of antipodal `hot spots' generated as a result of the thermonuclear 
burning of matter accreted on the neutron star surface. Indisputable evidence 
that neutron stars in X-ray binaries can indeed be rotating rapidly then came 
with the discovery of the first accreting millisecond X-ray pulsar SAX 
J1808-369 (Wijnands \& van der Klis 1998; Chakrabarty \& Morgan 1998), which 
is spinning at 401 Hz.  
\par The observed distribution of burst oscillation
frequencies and the 401 Hz pulsar is very similar to the observed distribution
of millisecond radio pulsars. The RXTE observations suggest that the spin 
frequencies of neutron stars in accreting binaries span a relatively narrow
range from $\sim 300 - 600$ Hz. Moreover, these observed frequencies are 
significantly less than the maximum neutron star spin rates for almost all but 
the stiffest neutron star equations of state (Cook, Shapiro \& Teukolsky 1994).
This has led to the suggestion that some mechanism may limit the spin periods 
of these accreting neutron stars. Bildsten (1998) recently proposed that the 
angular momentum gain from accretion could be offset by gravitational radiation
losses if a misaligned quadrupole moment of order $10^{-8} - 10^{-7} \; I_{NS}$
could be sustained in the neutron star crust. Here $I_{NS}$ is the moment of 
inertia of the neutron star. In this scenario the strong spin 
frequency dependence of the gravitational radiation losses sets the limit on 
the observed spin frequencies.  Recent theoretical work has also shown that 
an $r$-mode instability in rotating neutron stars may also be important in 
limiting the spins of neutron stars via gravity wave emission (see 
Ushomirsky, Bildsten \& Cutler 2000; Bildsten 
1998; Levin 1999; Andersson, Kokkotas, \& Stergioulas 1999).  
\par In the remainder of this contribution I will present an overview of the 
RXTE data and lay out the evidence for the conclusion that the observed burst 
oscillation frequencies are a manifestation of the spin frequencies of neutron 
stars (or perhaps twice the spin frquency in a few cases). I will summarize the
status of current X-ray measurements of neutron star spin periods and the 
implications for gravity wave emission. 

\section*{Burst Oscillations: Observational Overview}

Burst oscillations with a frequency of 363 Hz were first discovered from the 
LMXB 4U 1728-34 by Strohmayer et al. (1996). Oscillations in 
an additional nine sources have since been reported, with four of these only
appearing in the last few months. The sources and their 
observed frequencies are given in Table 1. In the remainder of this section I 
will briefly review the important observational properties of these 
oscillations and summarize the evidence supporting spin modulation as the 
mechanism. Because of the rapid pace of developments this will no doubt be an
incomplete review.

\subsection*{Oscillation Amplitudes near Burst Onset}

Some bursts show large amplitude oscillations during the $\approx 1 - 2$ s 
rises commonly seen in thermonuclear bursts. An example of this behavior in a 
burst from 4U 1636-53 is shown in figure 1. Strohmayer, Zhang \& Swank (1997)
showed that some bursts from 4U 1728-34 have oscillation amplitudes as large as
43\% within 0.1 s of the observed onset of the burst. They also showed that 
the oscillation amplitude decreased monotonically as the burst flux increased 
during the rising portion of the burst lightcurve. Strohmayer et al (1998a)
reported on strong pulsations in 4U 1636-53 at 580 Hz with an amplitude of 
$\approx 75$\% only $\sim 0.1$ s after detection of burst onset.
The presence of modulations of the thermal burst flux which can 
approach nearly 100\% right at burst onset fits nicely with the idea that 
early in the burst there exists a localized hot spot which is then modulated by
the spin of the neutron star. In this scenario the largest modulation 
amplitudes should be produced when the spot is smallest, and as the spot grows
to encompass more of the neutron star surface, the amplitude decreases. 
This behavior is consistent with the observations. X-ray spectroscopy during 
burst rise also indicates that the X-ray emission is localized on the neutron 
star near the onset of bursts. Strohmayer, Zhang, \& Swank (1997) found that 
during burst rise the flux is {\it underluminous} compared with intervals later
in the burst which have the same observed black body temperature, and suggested
that during the rise a localized, but growing segment of the surface of the 
neutron star is producing the X-ray emission. As the burst progresses the 
burning area increases in size until the entire surface is involved. 

\begin{figure}[b!] 
\centerline{\epsfig{file=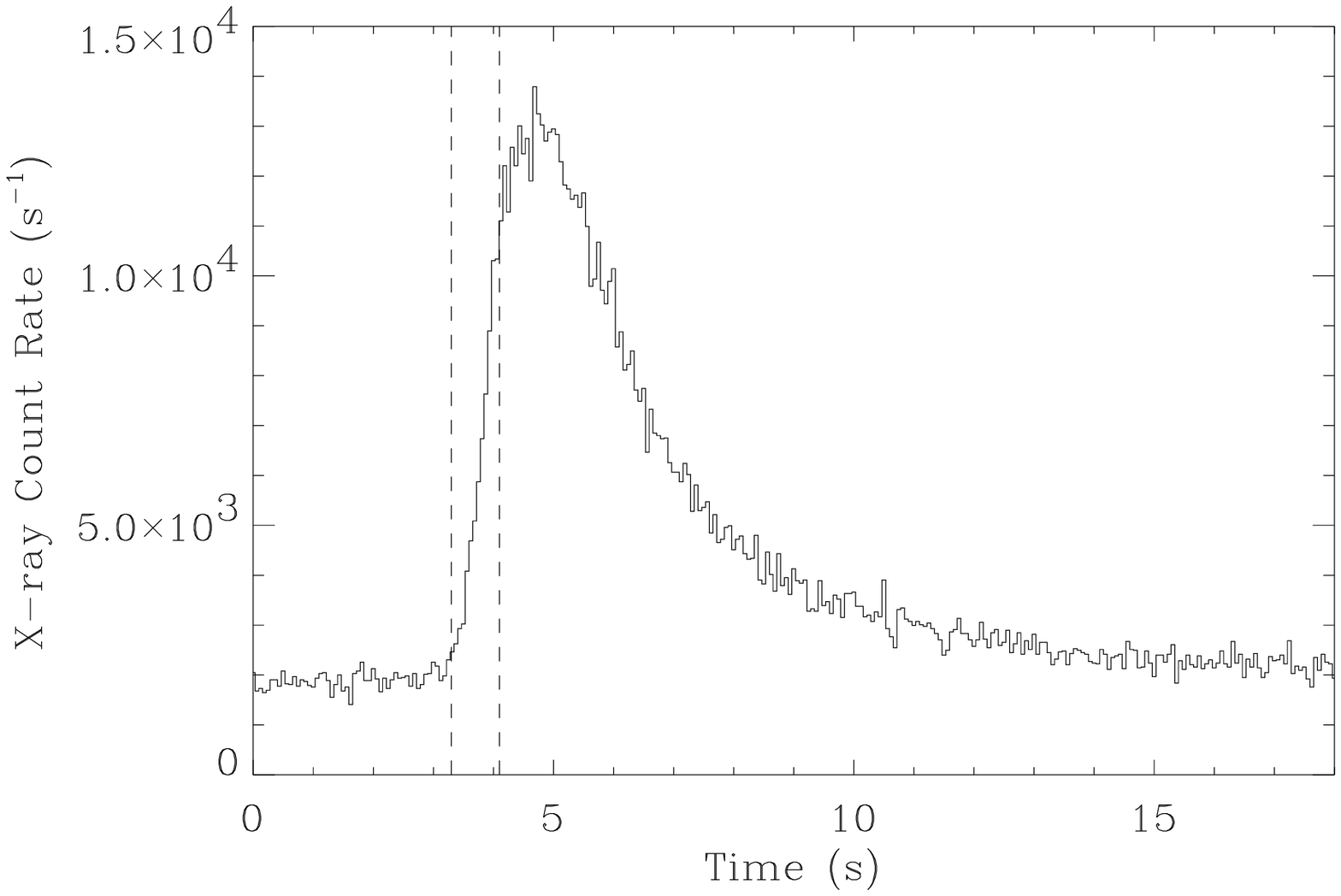,height=2.5in,width=2.9in}\epsfig
{file=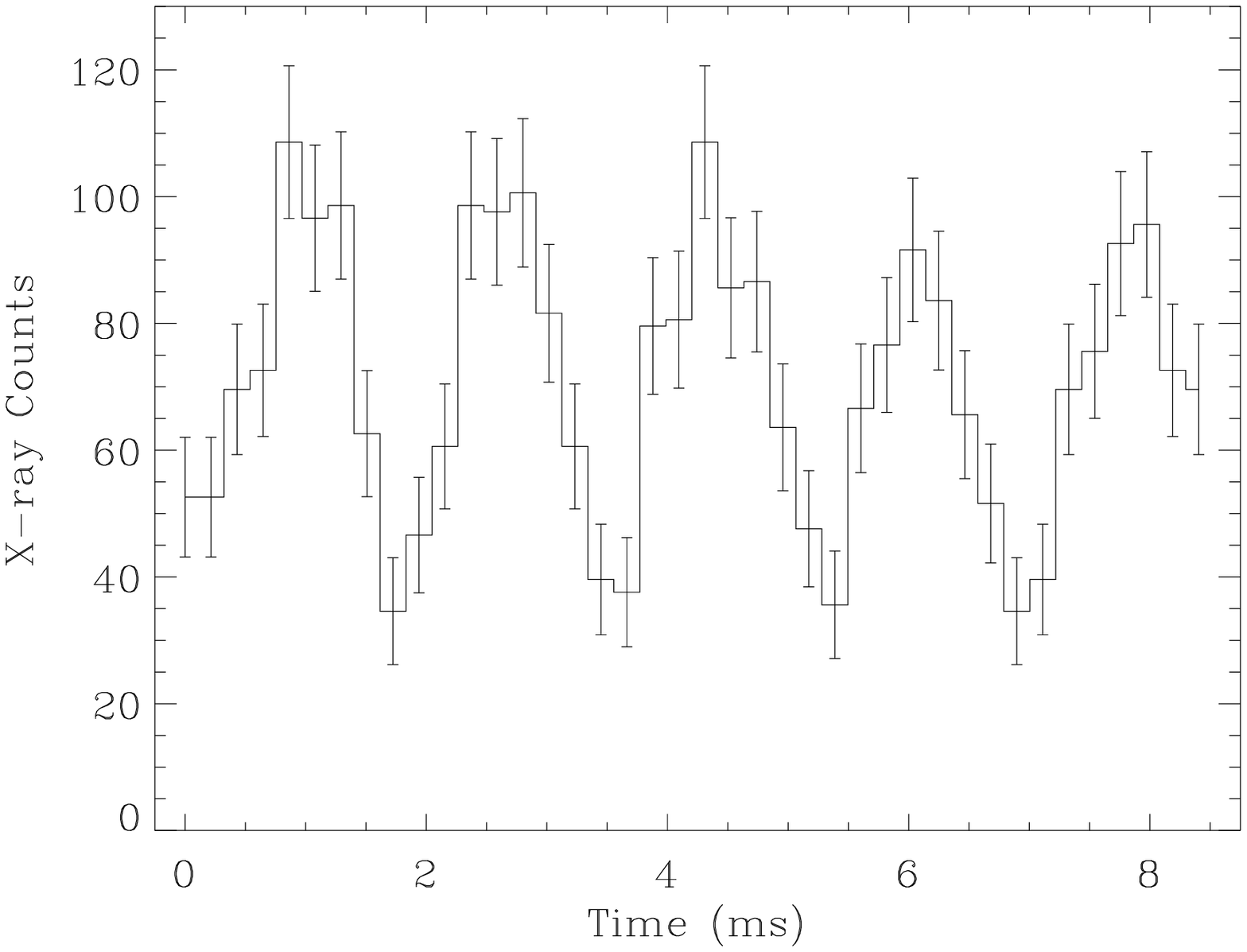,height=2.5in,width=2.9in}}
\caption{Burst oscillations during the rising phase of a burst from 4U 1636-53
at 580 Hz. The left panel shows the RXTE/PCA (2-60 keV) X-ray countrate during
a burst recorded on August 20th, 1998 at 05:14:09 UTC. The right panel shows 
the pulsations during the rising interval denoted by vertical dashed lines in 
the top panel. The date were folded at intervals of 5 $\times P_{spin}$ where 
$P_{spin} = 1.725$ ms. Note the large amplitude of the pulsations. The 
preburst countrate intensity of about 35 cts/sec has been subtracted.}
\label{fig1}
\end{figure}

\begin{table}[tbp]
\centerline{{\bf Table 1.} Burst Oscillation Sources and Properties}
\begin{tabular}{ccccccc}
\hline
 & & & Sources & Frequency (Hz) & $\Delta\nu$ (kHz QPO, in Hz) & 
References$^1$ \cr
\hline
 & & & 4U 1728-34 & 363 & 363 - 280 & 1, 2, 3, 4, 5, 13, 14 \cr
 & & & 4U 1636-53 & 290, 580 & 251 & 6, 7 \cr
 & & & 4U 1702-429 & 330 & 315 - 344 & 4, 9 \cr
 & & & KS 1731-260 & 524 & 260 & 10, 11, 12 \cr
 & & & Galactic Center & 589 & Unknown & 15 \cr
 & & & Aql X-1 & 549 & Unknown & 16, 17 \cr
 & & & X1658-298 & 567 & Unknown & 18 \cr
 & & & 4U 1916-053 & 270 & 290 - 348 & 19, 20 \cr
 & & & 4U 1608-52 & 619 & 225 - 325 & 8, 21 \cr
 & & & SAX J1808-369 & 401 & Unknown & 22,23 \cr
\hline
\end{tabular}
\hfil\hspace{\fill}

{\small $^1$References: (1) Strohmayer et al. (1996); (2) Strohmayer,
Zhang, \& Swank (1997); (3) Mendez \& van der Klis (1999); (4) Strohmayer 
\& Markwardt (1999); (5) Strohmayer et al. (1998b); (6) Strohmayer et al. 
(1998a); (7) Miller (1999); (8) Mendez et al. (1998); (9) Markwardt, 
Strohmayer
\& Swank (1999) (10) Smith, Morgan, \& Bradt (1997); (11) Wijnands \& van der
Klis (1997); (12) Muno et al. (2000); (13) van Straaten et al. (2000); (14)
Franco (2000); (15) Strohmayer et al (1997); (16) Zhang et al. (1998); (17)
Ford (1999); (18) Wijnands, Strohmayer \& Franco (2000); (19) Boirin et al. 
(2000); (20) Galloway et al. (2000); (21) Chakrabarty (2000); (22) Heise 
(2000); (23) Ford (2000)}
\end{table} 

\subsection*{Coherence and Stability of Burst Oscillations}

The observed oscillation frequency during a burst is usually not constant.
Often the frequency is observed to increase by $\approx 1 - 3$ Hz in the 
cooling tail, reaching a plateau or asymptotic limit (see Strohmayer et al. 
1998b). This behavior is common to all the burst oscillation sources, and it
would appear that the same physical mechanism is involved, however, there
have been reports of decreases in the oscillation frequency with time. For
example, Strohmayer (1999) and Miller (2000) identified a burst from 4U 1636-53
with a spin down of the oscillations in the decaying tail. This burst also had
an unusually long decaying tail which may have been related to the spin down
episode. Muno et al. (2000) reported an episode of spin down in a burst from
KS 1731-24. 
\par Strohmayer et. al (1997) have suggested that the time evolution of the 
burst oscillation frequency results from angular momentum conservation of the 
thermonuclear shell. The burst expands the shell, increasing its rotational 
moment of inertia and slowing its spin rate. Near burst onset the shell is 
thickest and thus the observed frequency lowest. The shell spins back up as it
cools and recouples to the underlying neutron star. Calculations indicate that 
the $\sim 10$ m thick pre-burst shell expands to $\sim 30-40$ m during the 
flash (see Joss 1978; Bildsten 1995; Cumming \& Bildsten 2000), which gives a 
frequency shift of $\approx 2 \ \nu_{spin} (20 \ {\rm m}/ R)$,
where $\nu_{spin}$ and $R$ are the stellar spin frequency and radius, 
respectively. For the several hundred Hz spin frequencies inferred from 
burst oscillations this gives a shift of $\sim 2$ Hz, similar to that observed.
Recently, Galloway et al. (2000) reported the discovery of a 270 Hz oscillation
in a burst from 4U 1608-52. They measured a frequency drift of $\sim 3.6$ Hz
during this burst and suggested that thermal expansion of the burning layer may
not be sufficient to explain the shift. We note that the current theoretical 
estimates do not include the rotational lowering of the effective gravity and
are also hydrostatic calculations. Dynamic motions of the layer may also 
contribute to changes in the height of the burning layer. These effects could
increase the height of the burning layer and allow for greater frequency 
drifts than have been currently calculated. 
\par Burst oscillations have a much higher coherence than is typical for other
quasiperiodic X-ray variations observed from neutron star systems. For example,
the kHz QPO seen in many LMXBs have maximum coherence values, $Q \equiv \nu
/\Delta\nu \sim 100$. Strohmayer \& Markwardt (1999) showed that the frequency 
evolution of burst oscillations in 4U 1728-34 and 4U 1702-429 is highly phase 
coherent. They modelled
the frequency drift and showed that a simple exponential ``chirp" model of the 
form $\nu (t) = \nu_0 (1 - \delta_{\nu} \exp(-t/\tau) )$, works remarkably 
well, producing quality factors $Q \equiv \nu_0 / \Delta\nu_{FWHM} \sim 4,000$.
Muno et al. (2000) performed a similar analysis on bursts from KS 1731-26 and
concluded that the burst oscillations from this source were also phase 
coherent. These results argue strongly that the mechanism which produces the 
modulations is intrinsically a highly coherent one. 
\par The accretion-induced rate of change of the neutron star spin frequency 
in a LMXB is approximately $1.8 \times 10^{-6}$ Hz yr$^{-1}$ for typical 
neutron star and LMXB parameters.  The Doppler shift due to orbital motion of 
the binary can produce a frequency shift of magnitude
$ \Delta\nu / \nu = v \sin i /c \approx  2.05 \times 10^{-3}$, again for
representative LMXB system parameters. This doppler shift easily dominates 
over any possible accretion-induced spin change on orbital to several year 
timescales. Therefore the extent to which the observed burst oscillation 
frequencies are consistent with possible orbital Doppler shifts, but 
otherwise stable over $\approx$ year timescales, provides strong support 
for a highly coherent mechanism which sets the observed frequency. 
At present, the best source available to study the long term stability of burst
oscillations is 4U 1728-34. Strohmayer et al. (1998b) compared the observed
asymptotic frequencies in the decaying tails of bursts separated in time
by $\approx 1.6$ years. They found the burst frequency to be highly stable,
with an estimated time scale to change the oscillation period of 
about 23,000 year. We illustrate this behavior in figure 2 which compares
the observed burst oscillation frequency of two bursts from 4U 1728-34 which 
ocurred $\sim 3$ years apart. van Straaten et al. (2000) showed evidence that 
the frequency track made by a given burst from 4U 1728-34 is dependent on the 
position in the X-ray color - color diagram (a surrogate for mass accretion
rate). The two bursts shown in figure 2 have similar frequency tracks and there
closeness in frequency over a span of 3 years argues for a highly stable
process, such as rotation, as the mechanism which sets the frequency. 
\par Strohmayer et al. (1998b) also suggested that the stability of the 
asymptotic periods might be used to infer the X-ray mass function of
LMXB by comparing the observed asymptotic period distribution of many
bursts and searching for an orbital Doppler shift. The source 4U 1636-53 is a 
good candidate for such an effort because its orbital period is known (3.8
hrs). Strohmayer et al. (1998b) compared the highest observed frequencies in 
three different bursts from 4U 1636-53. The frequencies in these bursts alone 
were consistent with a typical orbital velocity for the neutron star. However, 
study of additional bursts reveals a greater range of highest 
frequencies than can likely be accounted for by orbital motion alone (see 
Giles \& Strohmayer 2001). A possible explanation of this within the context of
the spin modulation scenario is that not every burst has relaxed to the 
asymptotic value before the oscillations fade below the detection level. 
Nevertheless, the observed distribution does suggest the existence of an upper 
limit, which can naturally be associated with the spin frequency (Giles \& 
Strohmayer 2001).

\begin{figure}[b!] 
\centerline{\epsfig{file=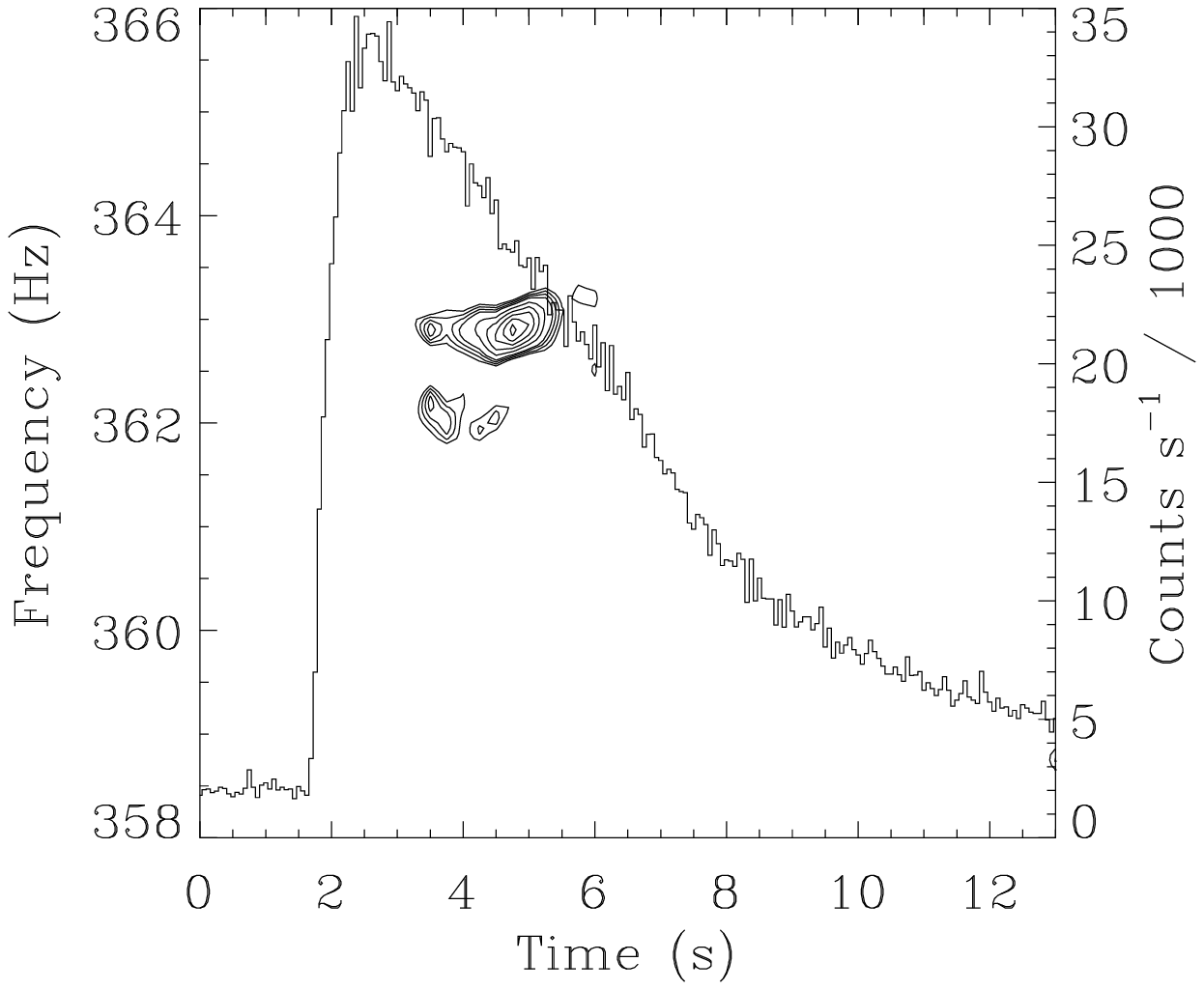,height=2.5in,width=2.95in}\epsfig
{file=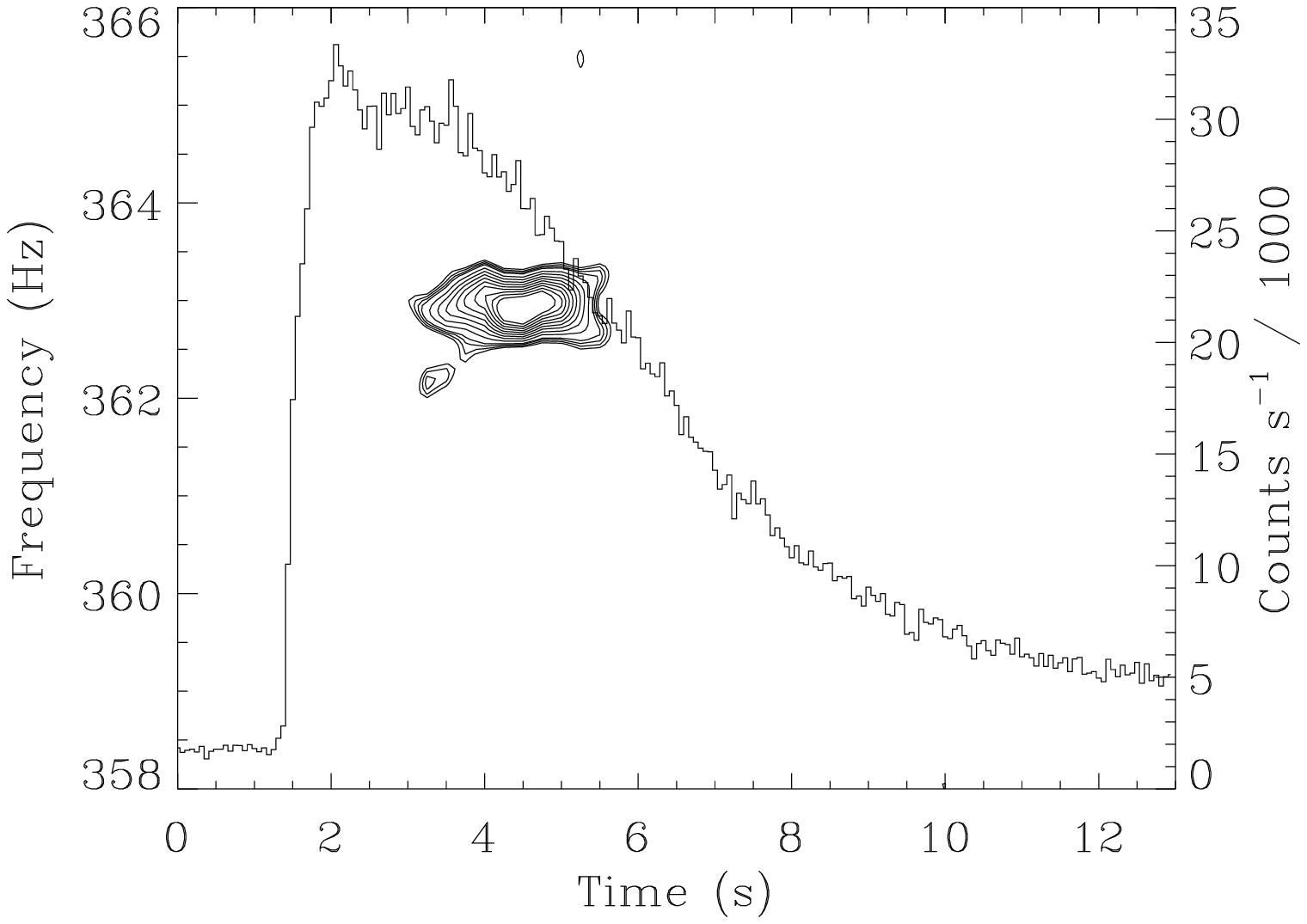,height=2.5in,width=2.95in}}
\caption{Burst oscillations in two bursts from 4U 1728-34 separated in time by 
$\sim 3$ yr. The frequency tracks are almost identical and suggests that the 
mechanism which sets the frequency is highly stable. The burst on the left was
observed in Feb. 1996 while that on the right was seen in Feb. 1999.}
\label{fig2}
\end{figure}

\subsection*{Burst Oscillations and Mass Accretion Rate}

Recent studies have focused on how the presence and properties of
burst oscillations correlate with other properties of these sources, for 
example, their spectral state and inferred mass accretion rates. 
Muno et al. (2000) were the first to conduct such a study and found that
bursts from KS 1731-26 with oscillations appear to only occur when the source
is on the banana branch in the X-ray color-color diagram. They also found that
these bursts were all photospheric radius expansion bursts. Cumming \& Bildsten
(2000) suggested that such bursts were likely pure Helium flashes and that it 
would be more likely for these to show oscillations because the radiative 
diffusion time is short compared to the inferred shearing time of the 
thermonuclear burning layer, making it more likely that a modulation would 
survive. Franco (2000) and van Straaten et al. (2000) showed that bursts from 
4U 1728-34 with oscillations also occur preferentially on the banana branch, 
but they did not find a similar relationship with radius expansion as for KS 
1731-26. They also found that the portion of a full frequency track which is 
present in a burst appears to also depend on mass accretion rate. Franco (2000)
also found that for 4U 1728-34 the strength of oscillations was correlated with
position in the color-color diagram. Since other burst properties, such as 
peak flux, fluence and durations, are also known to correlate with mass 
accretion rate it is not surprising, given the fact that the physics of 
thermonuclear burning is dependent on mass accretion rate, that the properties 
of burst oscillations appear also to be strongly dependent on mass accretion 
rate. 

\subsection*{Theoretical Expectations}

\par The notion that X-ray bursts are caused by thermonuclear instabilities in 
the accreted layer on the surface of a neutron star is now universally 
accepted. There is no doubt that interesting puzzles remain and our detailed 
understanding is incomplete, but the basic model is firmly established.
The thermonuclear instability which triggers an X-ray burst burns in a few 
seconds the fuel which has been accumulated on the surface 
over several hours. This makes it very unlikely that the conditions 
required to trigger the instability will be achieved simultaneously over the 
entire stellar surface. This notion, first emphasized by Joss (1978), 
led to the study of lateral propagation of the burning front over the neutron 
star surface (see Fryxell \& Woosley 1982, Nozakura, Ikeuchi \& Fujimoto 1984, 
and Bildsten 1995). The short risetimes of thermonuclear X-ray bursts 
suggest that convection plays an important role in the physics of the burning 
front propagation, especially in the low accretion rate regime which leads to 
large ignition columns (see Bildsten 1998 for a recent review of
thermonuclear burning on neutron stars). These 
studies emphasized that the physics of thermonuclear burning is necessarily a 
multi-dimensional problem and that {\it localized} burning is to be expected, 
especially at the onset of bursts.  The properties of oscillations near burst 
onset described above fit nicely with this picture of thermonuclear burning on 
neutron stars.

\section*{Implications for Gravity Wave Emission}

The strong $\nu_s^6$ frequency dependence of the energy radiated by 
gravitational waves means that rapidly rotating neutron stars with misaligned 
quadrupole moments might have observationally interesting gravitational wave 
amplitudes. The spin periods of neutron stars inferred from burst oscillations 
cluster rather tightly in the range from $\sim 300 - 600$ Hz. As pointed out 
earlier, these frequencies are well below the maximum break-up frequencies for 
most modern neutron star equations of state (Cook, Shapiro \& Teukolsky 1994). 
White \& Zhang (1997) suggested that the observed range of spin frequencies
could be produced if these neutron stars were spinning in magnetic equilibrium.
However, in order for the observed frequencies to be similar would require that
$\dot M$ and the magnetic moment $\mu_{b}$ be correlated. It is not presently 
known if such a correlation is to be naturally expected based on theoretical 
grounds. 

\subsection*{Crustal Deformation Quadrupole Moments}

\par An alternative model has been proposed by Bildsten (1998). He suggested 
that the spins of these neutron stars may be limited by the emission of gravity
waves. The spin down torque due to gravitational wave emission is proportional
to $\nu_s^5$ so that one would expect a critical spin 
frequency above which accretion torques are cancelled out by gravity wave 
losses. By equating the characteristic accretion torque with the gravity wave
torque one can determine the average quadrupole moment required to maintain the
critical frequency. For the mass accretion rates characteristic of LMXBs and
a critical spin frequency of 300 Hz one obtains a quadrupole $Q \sim 
4.5 \times 10^{37}$ g cm$^2$, or about $5\times 10^{-8} \; I_{NS}$ (see 
Bildsten 1998). The question that remains is whether or not such a quadrupole
moment can be routinely generated in a neutron star. 
\par The idea that the spin frequencies of neutron stars might be limited by 
gravitational radiation losses was initially proposed by Papaloizou \& Pringle 
(1978) and Wagoner (1984). Wagoner (1984)  
argued that the Chandrasekhar - Friedman - Schutz instability would excite 
non-axisymmetric modes which would radiate gravity waves and limit the spin 
frequency. However, Lindblom (1995) and Lindblom \& Mendell (1995) showed that 
this instability will only set in near the break-up frequency, which is at
much higher frequency than the observed burst oscillations for most modern
equations of state. Bildsten (1998) has suggested that electron captures on
heavy nuclei in the neutron star crust might be able to produce a quadrupole of
the required amount. The basic idea is that the electron capture process
produces density jumps in the crust. Since the electron capture
rate is temperature sensitive in the crust a lateral temperature gradient 
could lead to density jumps which as a function of lateral position occur at 
slightly different depths in the crust. 
Ushomirsky, Cutler, \& Bildsten (2000) have investigated
this process in more detail and concluded that electron captures could produce
the observed quadrupole if there are $\sim 5 \%$ lateral temperature
variations at crustal depths where the density is in excess of $10^{12}$ 
g cm$^{-3}$. They also 
computed the dimensionless strain $\sigma$ which gives a quadrupole sufficient 
to balance the accretion torque and found that $\sigma \sim 10^{-2}$ at near
Eddington accretion rates. Although promising, more theoretical work will be 
required to convincingly establish whether this mechanism can produce a 
sufficient quadrupole moment to balance the accretion torque. 
\par Regardless of the mechanism, if the accretion torque is indeed balanced by
gravity wave losses then the amplitude of the gravitational radiation can be
calculated (see Wagoner 1984; Bildsten 1998). The dimensionless strain $h$ 
is in the range from $h \sim 10^{-27}  - 10^{-26}$. Although this strain is 
significantly less than the estimated sensitivity for LIGO I, one can greatly
improve the sensitivity by pulse folding if the rotational ephemeris of the 
neutron star is known (see Brady \& Creighton 1999; Ushomirsky, Bildsten \& 
Cutler 2000). Current estimates indicate that a narrow band configuration for 
LIGO-II will reach interesting search limits for these neutron stars, 
especially for the brightest of the LMXBs, for example, Sco X-1 (Ushomirsky, 
Bildsten \& Cutler 2000). This also provides strong motivation for additional
deep X-ray timing searches in order to detect coherent pulsations in more 
LMXBs and to measure the pulse ephemerides so as to improve searches for 
gravity wave emission. It also illustrates the strong synergism between 
X-ray and gravity wave astronomy in the context of neutron star binaries.

\subsection*{R-Mode Instability}

Andersson (1998) recently discovered that the $r$-modes of rotating 
relativistic stars are excited by gravitational radiation at all rotation
frequencies (see also Friedman \& Morsink 1998). Shear and bulk viscosity 
damps these modes, so whether or not they can attain significant amplitudes 
depends on the competition between gravitational radiation excitation and 
viscous damping. This discovery has led to a flurry of theoretical activity to 
try and understand the implications of the $r$-mode instability for rapidly 
rotating neutron stars. Here I will only give a brief summary. See the 
contribution by Ushomirsky (2001, these proceedings) for all the details.
\par For accreting neutron stars the basic idea is that a star will be spun 
up by accretion to some critical frequency at which the $r$-mode instability
sets in. Work by Andersson, Kokkotas \& Stergioulas (1999) suggested that the
star would spin up to this critical frequency and then remain in equilibrium
at this frequency with the accretion torque balanced by angular momentum losses
from the excited $r$-modes. However, Levin (1999) and Spruit (1999) showed that
this evolution is not possible because the $r$-modes heat the stellar 
interior which reduces the viscosity and increases the growth rate. Thus the 
$r$-modes would grow rapidly and spin the star down on a very short timescale
($\sim$ a few months), much shorter in fact than the timescale to spin the 
star up via accretion. During this time the star would be a powerful source of
gravity waves, however, the timescale is so short that the effective event
rate is very low (see Levin 1999; Andersson et al. 2000). 
\par Recent efforts have
focused on studying the sources of viscous damping, most importantly the 
influence of the solid crust of the neutron star. Bildsten \& Ushormirsky 
(2000) estimated the effect the solid crust would have on the damping and 
concluded the crust disspation would greatly exceed that from standard 
viscosity in the core. More recently, Andersson et al. (2000) have revised 
these estimates and argue that the damping is not as strong as suggested by 
Bildsten \& Ushormirky (2000). They conclude that the $r$-mode instability 
could explain the observations of most millisecond pulsars with periods between
about 1.5 and 6 ms as well as the lack of any pulsars spinning faster than 
$\sim 1.5$ ms. 

\section*{Acknowledgements}

\par I would like to thank the organizers of the Astrophysical Sources of 
Gravitational Radiation Meeting for their gracious hospitality and for 
producing such a stimulating scientific program. I thank Jean Swank, 
Greg Ushomirsky, and Lars Bildsten for many helpful discussions.


\begin{references}
\bibitem{And00}Andersson, N., Jones, D. I., Kokkotas, K. D. \& Stergioulas, N.
2000, ApJ, L75
\bibitem{Andersson et al.}Andersson, N. Kokkotas, K. D. \& Stergioulas, N. 
1999, ApJ, 516, 307
\bibitem{And}Andersson, N. 1998, ApJ, 502, 708
\bibitem{Barish}Barish, B. 2001, these proceedings
\bibitem{BU00}Bildsten, L. \& Ushomirsky, G. 2000, ApJ, 529, L33
\bibitem{B98}Bildsten, L. 1998, in ``The Many Faces of Neutron Stars'', ed. R. 
Buccheri, A. Alpar \& J. van Paradijs (Dordrecht: Kluwer), p. 419
\bibitem{B98ApJ}Bildsten, L. 1998, ApJ, 501, L89
\bibitem{B95}Bildsten, L. 1995, ApJ, 438, 852
\bibitem{Boirin}Boirin, L., et al. 2000, A\& A, 361,121
\bibitem{Brady}Brady, P. \& Creighton, T. 1999, Phys. Rev. D, in press
(gr-qc/9812014)
\bibitem{Chaky00}Chakrabarty, D. 2000, Talk presented at AAS HEAD meeting, 
Honolulu, HI
\bibitem{CM}Chakrabarty, D. \& Morgan, E. H. 1998, Nature, 394, 346
\bibitem{cs}Chen, K. \& Shaham, J. 1989, ApJ, 339, 279
\bibitem{CST}Cook, G. B., Shapiro, S. L. \& Teukolsky, S. A. 1994, ApJ, 424,
823
\bibitem{CB00}Cumming, A. \& Bildsten, L. 2000, ApJ, in press 
(astro-ph/0004347)
\bibitem{faber}Faber, J. \& Rasio, F. 2001, these proceedings
\bibitem{Ford}Ford, E. C. 2000, ApJ, 535, L119
\bibitem{Ford2}Ford, E. C. 1999, ApJ, 519, L73
\bibitem{Franco}Franco, L. 2000, ApJ, submitted (astro-ph/0009189)
\bibitem{FM98}Friedman, J. L. \& Morsink, S. 1998, ApJ, 502, 714
\bibitem{FW}Fryxell, B. A., \& Woosley, S. E. 1982, ApJ, 261, 332
\bibitem{G00}Galloway et al. 2000, ApJ, submitted (astro-ph/0010072)
\bibitem{GS00}Giles, A. B. \& Strohmayer, T. E. 2001, in preparation
\bibitem{Heise}Heise, J. et al. 2000, Talk presented at AAS HEAD meeting, 
Honolulu, HI
\bibitem{1999ApJ...525L..45H}Heiselberg, H.\ \& Hjorth-Jensen, M.\ 1999, ApJ, 
525, L45
\bibitem{Joss}Joss, P. C. 1978, ApJ, 225, L123
\bibitem{Lee}Lee, W. 2001, these proceedings
\bibitem{Levin}Levin, Y. 1999, ApJ, 517, 328
\bibitem{Lind}Lindblom, L. 1995, ApJ, 438, 265
\bibitem{LindMen}Lindblom, L. \& Mendell, G. 1995, ApJ, 444, 804
\bibitem{mss}Markwardt, C. B., Strohmayer, T. E., \& Swank, J. H. 1999, ApJ, 
512, L125
\bibitem{Men et al.}Mendez, M., van der Klis, M. \& van Paradijs, J. 1998, 
ApJ, 506, L117
\bibitem{Mendez}Mendez, M. \& van der Klis, M. 1999, ApJ, 517, L51
\bibitem{2000ApJ...531..458M}Miller, M.\ C.\ 2000, ApJ, 531, 458
\bibitem{m}Miller, M. C. 1999, ApJ, 515, L77
\bibitem{2000ApJ...542.1016M}Muno, M.\ P., Fox, D.\ W., Morgan, E.\ H.\ \& 
Bildsten, L.\ 2000, ApJ, 542, 1016
\bibitem{NIF}Nozukura, T., Ikeuchi, S., \& Fujimoto, M. Y. 1984, ApJ, 286, 221
\bibitem{Papaloizou}Papaloizou, J. \& Pringle, J. E. 1978, MNRAS, 184, 501
\bibitem{SMB}Smith, D., Morgan, E. H. \& Bradt, H. V. 1997, ApJ, 479, L137
\bibitem{Spruit}Spruit, H. C. 1999, A\& A, 341, L1
\bibitem{stroh00}Strohmayer, T. E. 2000, in Proceedings of X-ray Astronomy '99,
Stellar Endpoints, AGN and the Diffuse X-ray Background. Bologna, Italy, 
(astro-ph/9911338)
\bibitem{sm}Strohmayer, T. E. \& Markwardt, C. B. 1999, ApJ, 516, L81
\bibitem{stroh99}Strohmayer, T. E. 1999, ApJ, 523, L51
\bibitem{s98a}Strohmayer, T. E., Zhang, W., Swank, J. H., White, N. E. \& 
Lapidus, I. 1998a, ApJ, 498, L135
\bibitem{s98b}Strohmayer, T. E., Zhang, W., Swank, J. H. \& Lapidus, I. 1998b, 
ApJ, 503, L147
\bibitem{SJGL}Strohmayer, T. E., Jahoda, K., Giles, A. B. \& Lee, U. 1997, 
ApJ, 486, 355
\bibitem{szs}Strohmayer, T. E., Zhang, W. \& Swank, J. H. 1997, ApJ, 487, L77
\bibitem{Stroh et al.}Strohmayer, T. E. et al. 1996, ApJ, 469, L9
\bibitem{Usho}Ushomirsky, G. 2001, these proceedings
\bibitem{UBC}Ushomirsky, G., Bildsten, L. \& Cutler, C. 2000, in 3rd Edoardo
Amaldi Conference on Gravitational Waves, (astro-ph/0001129)
\bibitem{UCB}Ushomirsky, G., Cutler, C. \& Bildsten, L. 2000, MNRAS, submitted,
(astro-ph/0001136)
\bibitem{1995A&A...297L..41V}van den Heuvel, E.\ P.\ J.\ \& Bitzaraki, O.\ 
1995, A\& A, 297, L41
\bibitem{1997A&A...317L...9V}van Paradijs, J., van den Heuvel, E.\ P.\ J., 
Kouveliotou, C., Fishman, G.\ J., Finger, M.\ H.\ \& Lewin, W.\ H.\ G.\ 1997, 
A\& A, 317, L9
\bibitem{vanStraaten}van Straaten, S. et al. 2000, ApJ, submitted, 
(astro-ph/0009194)
\bibitem{Wagoner}Wagoner, R. V. 1984, ApJ, 278, 345
\bibitem{WRS}Webbink, R. F., Rappaport, S. A. \& Savonije, G. J. 1983, ApJ, 
270, 678
\bibitem{WZ}White, N. E. \& Zhang, W. 1997, ApJ, 490, L87
\bibitem{WSF}Wijnands, R. Strohmayer, T. E. \& Franco, L. M. 2000, ApJ, in 
press (astro-ph/0008526)
\bibitem{WvdK}Wijnands, R. \& van der Klis, M. 1998, Nature, 394, 344
\bibitem{Wvdk2}Wijnands, R., \& van der Klis, M. 1997, ApJ, 482, L65
\bibitem{Zet al.}Zhang, W. et al. 1998, ApJ, 495, L9-12

\end{references}
\end{document}